\documentclass[sigconf]{acmart}

\copyrightyear{2026}
\acmYear{2026}
\setcopyright{cc}
\setcctype{by}
\acmConference[CODASPY '26] {Proceedings of the Sixteenth ACM Conference on Data and Application Security and Privacy}{June 23--25, 2026}{Frankfurt am Main, Germany.}
\acmBooktitle{Proceedings of the Sixteenth ACM Conference on Data and Application Security and Privacy (CODASPY '26), June 23--25, 2026, Frankfurt am Main, Germany}
\acmDOI{10.1145/3800506.3803491}
\acmISBN{979-8-4007-2562-3/2026/06}

\usepackage{
xcolor}

\usepackage{soul}

\usepackage{tfrupee}
\newcommand{\INR}[1]{\rupee~#1}

\usepackage{etoolbox}
\makeatletter
\patchcmd\@thm
 {\let\thm@indent\noindent}%
  {}{}
\makeatother

\usepackage{dsfont}
\usepackage{enumitem}

\usepackage{booktabs} 
\usepackage{mdframed}
\usepackage{framed}

\usepackage{hyphenat}
\usepackage{fnpct}

\usepackage{algpseudocode}
\usepackage{amsmath,epsfig}
\usepackage[boxed,ruled,linesnumbered, longend, noend]{algorithm2e}
\usepackage{amsthm}
\usepackage{setspace}
\usepackage{enumitem}
\usepackage{multirow}
\usepackage{hhline}
\usepackage{caption}
\usepackage{subcaption}
\usepackage{graphicx}
\usepackage{wrapfig}
\usepackage{amsmath,epsfig}
\usepackage{amsthm}

\usepackage{tikz}
\usepackage{balance}

\usepackage{enumitem}
\usepackage{tablefootnote}
\usepackage{fnpct}
\usepackage{setspace}

\makeatletter
\def\thm@space@setup{\thm@preskip=0pt
\thm@postskip=0pt}
\makeatother

\usepackage{array}
\usepackage{makecell}

\newcommand{\B}{\vspace*{-\smallskipamount}}
\newcommand{\BB}{\vspace*{-\medskipamount}}
\newcommand{\BBB}{\vspace*{-\bigskipamount}}

\usepackage{longtable}
\usepackage{mdframed}
\usepackage{colortbl}

\usepackage[hang,flushmargin]{footmisc}

\usepackage{natbib}
\newcommand*\wrapletters[1]{\wr@pletters#1\@nil}
\def\wr@pletters#1#2\@nil{#1\allowbreak\if&#2&\else\wr@pletters#2\@nil\fi}
\usepackage{tikz}
\usetikzlibrary{automata,matrix,shapes,arrows,positioning,chains,calc}
\usetikzlibrary{snakes}
\usetikzlibrary{arrows,scopes}
\usetikzlibrary{positioning,chains,fit,shapes,calc}
\usepackage{caption}
\usepackage{subcaption}
\usepackage{amsmath}
\usepackage{lineno}
\usepackage{pifont}

\definecolor{lightyellow}{HTML}{ABEBC6}
\sethlcolor{lightyellow}

\definecolor{myblue}{HTML}
{BF40BF}

\mdfdefinestyle{MyFrame}{%
    linecolor=white,
    outerlinewidth=2pt,
    innertopmargin=4pt,
    innerbottommargin=4pt,
    innerrightmargin=4pt,
    innerleftmargin=4pt,
        leftmargin = 4pt,
        rightmargin = 4pt,
        backgroundcolor=yellow!41!white}

\begin{document}

\ccsdesc[500]{Security and privacy~Privacy protections}
\ccsdesc[500]{Security and privacy~Social aspects of security and privacy}
\ccsdesc[500]{Security and privacy~Usability in security and privacy}

\keywords{Data protection laws; user privacy}

\onecolumn

\title{Local Privacy Laws in a Globalized World}



\author{Shantanu Sharma}
\affiliation{
  \institution{New Jersey Institute of Technology}
  \city{Newark}
  \country{USA}
}

\author{Ethan Myers}
\affiliation{
  \institution{Colorado State University}
  \city{Fort Collins}
  \country{USA}
}

\author{Lorenzo De Carli}
\affiliation{
  \institution{University of Calgary}
  \city{Calgary}
  \country{Canada}
}

\author{Ritwik Banerjee}
\affiliation{
  \institution{Stony Brook University}
  \city{Stony Brook}
  \country{USA}
}

\author{Indrakshi Ray}
\affiliation{
  \institution{Colorado State University}
  \city{Fort Collins}
  \country{USA}
}




\begin{abstract}
    Personal data has emerged as a highly valuable yet sensitive asset that drives business decisions, enables targeted advertising, and generates substantial revenue for companies, while simultaneously facilitating invasive monitoring of users. In recent years, research on digital privacy violations, including undue access, collection, and sharing of user data, has grown significantly. Much of this research adopts the European General Data Protection Regulation (GDPR) as the primary reference framework. 
    This is reasonable, as GDPR was a pioneering legislation, and many of its stipulations are clear and unambiguous. However, we argue that focusing solely on GDPR (and a small set of other Western regulatory frameworks) ignores privacy-related concerns, attitudes, and problems faced by users from other locales, creating a significant research blind spot.


This work systematically normalizes the heterogeneous legal requirements of multiple data protection laws into a unified abstraction aligned with the data lifecycle, which forms the foundation for the implementation of such regulations. We further investigate the implications of these laws on different stakeholders, including users, organizations, and governments. Overall, this work aims to broaden the digital privacy research community's perspective and to serve as a set of guiding principles for developing technological privacy solutions spanning multiple countries.


\end{abstract}

\maketitle

{

\section{Introduction}
\label{sec:intro}
\BB

Personal data is a valuable and sensitive asset, driving business decisions, enabling targeted advertising, and generating revenue for companies. However, this potential for monetization also creates significant privacy risks. Investigations in data management practices have repeatedly found issues, e.g.,  overcollection~\cite{zimmeck_automated_2017, reyes_wont_2018, liu_opted_2024, kollnig_fait_nodate}, undue data sharing~\cite{nguyen_share_nodate, liu_opted_2024, fan_empirical_2020,  Reyes2017IsOC}, and negligent data management~\cite{ihunter_2024, jia_who_2019, schmidt_2025}. Probes of user attitudes suggest that users may underestimate the risk of such violations~\cite{Boutet_Morel_2025, Dhondt_294510_2024, zufferey_revoked_2023, 10.5555/3563572.3563580, 10.5555/3361476.3361480}, or misunderstand the implications of data collection~\cite{xiang_policychecker_2023, qamar_detecting_2021, nguyen_freely_2022, mohan_analyzing_2019, koch_ok_nodate}. Given these findings, it is important for academic research to continue shedding light on issues related to digital privacy.


Privacy research typically defines a \textit{privacy threat model}, outlining specifically how data collection/processing operations may create risks for the owner of the data. 
While in some work this model is defined implicitly or based on informal considerations (\cite{ardalani2024, pandita_whyper_2013, feng_acnet_2019, feichtner_2020, qu_autocog_2014, neptune_2022, abhinav_2021}), the majority of research effort measure privacy violations with respect to a regulatory framework.\footnote{\scriptsize \textit{E}.\textit{g}.,~\cite{Reyes2017IsOC, reyes_wont_2018, nguyen_freely_2022, nguyen_share_nodate, fan_empirical_2020, xiang_policychecker_2023, liu_opted_2024, sassetti_assurance_2023, li_are_2024, zimmeck_automated_2017, mohan_analyzing_2019, ferreira_rulekeeper_2023, zhang-kennedy_whether_nodate, qamar_detecting_2021, singh_technical_2020, herwanto_leveraging_2024} and see {\color{blue}Table ~\ref{tab:privacy_studies}}.}
Doing so is reasonable, as regulations define unambiguous expectations that reflect the expectations of relevant stakeholders (government, citizens) in a given locale. 
Unfortunately, the current body of research does not reflect the global diversity of \emph{\textbf{data protection privacy laws}}. 
An exploratory review of 21 recent digital privacy-focused papers (see {\color{blue}Table~\ref{tab:privacy_studies}}) shows that a simple majority of them (14 out of 21) only consider the European General Data Protection Regulation (GDPR); and more than 95\% (20 out of 21) only consider European and/or North American regulatory frameworks (GDPR/CCPA/COPPA/CalOPPA/PIPEDA) 
--- also refer to {\color{blue}\S\ref{subsec:Limitations of Using Localized Laws}} to see our discussion on how existing work focused on specific definitions of personal/user data and consent models, and hence introducing a blind spot in global application of data protection laws.
This narrow geographic focus opens up questions about generalizability; as it may overlook privacy-related concerns, attitudes, and problems of non-European and more broadly non-Western users, causing a significant research blind spot.

As privacy researchers, we believe it is important for research in this space to incorporate awareness of the digital privacy concerns of the global user community. 
Thus, this work aims to provide a comparative analysis of data protection laws/regulations, considering frameworks beyond North America and Europe, including those from Asia, Africa, and South America. 
Specifically, we review  GDPR (EU), CCPA (California, US), DPDPA (India), PIPL (China), POPIA (South Africa), and LGPD (Brazil).\footnote{\scriptsize
{\color{black}
A comparative analysis of every possible data protection regulation is beyond the scope of a single paper, and we consider it unnecessary. Instead, we argue that it is enough to show structural differences between the GDPR and a small number of selected other regulations as counterexamples.}}

\smallskip\noindent
\textbf{Our objective.}
\emph{{Our objective is to examine how different jurisdictions implement data protection through regulatory laws.}} 
{\color{black}
While our goal is not to achieve exhaustive global coverage, we identify structurally distinct regulatory paradigms and demonstrate that overlooking these differences introduces critical blind spots in privacy research.
}
This paper provides researchers with knowledge, ensuring the regulatory foundations of technical work on privacy are aware of and informed by the global diversity of regulations. Overall, the data protection laws considered in this paper apply to $\approx$38.3\% of the world population.


\smallskip
\noindent
\textbf{Data protection laws and data life cycle.}
Conducting such a review is challenging for several reasons, such as
(i)~regulations across diverse countries are shaped by distinct regulatory traditions and practices, as well as by different conceptualizations of privacy rights and violations;
(ii)~the scope of the law itself may differ, with regulations in different locales targeting different aspects of the privacy problem;
(iii)~differences in how they treat user rights for actions such as data breaches, unauthorized data collection, and unauthorized data sharing.

To systematically understand these regulations, thus, we analyze them through the lens of the digital data processing life cycle, {\color{black} which serves as the operational axis, where users, organizations, and governments are not independent entities; rather, they represent different levels at which privacy is operationalized. We demonstrate how these laws influence each stage of the data life cycle and identify potential threats that may compromise privacy at these stages. Specifically, 
we study data protection laws' differences among entities and their roles based on data lifecycle abstraction, such as collection, processing, storage,  transfer, and deletion, as shown in {\color{blue}Table~\ref{tab:data_lifecycle_mapping}}.}

Overall, we find that data protection laws differ significantly in scope, definitions, and enforcement mechanisms. 
This creates challenges for both researchers who intend to measure and/or address privacy issues. 
Overall, we tackle the following research questions:

\medskip
{\bf \noindent \ding{228} RQ1: How do differences in user-level legal protections lead to inconsistent legal outcomes for the same digital data-collection behavior?} 

{\color{blue}\S\ref{subsec:Laws in terms of the Users}} shows that different laws define personal and sensitive data differently and also differ significantly in the rights users can exercise to control their data. 


\medskip
{\bf \noindent \ding{228} RQ2: How do differences in organizational obligations across data protection laws constrain the design of a unified and technically enforceable data-processing architecture?} 

Organizations with a global footprint need to implement data-processing principles, breach-notification procedures, cross-border transfer rules, deletion requirements, and accountability mechanisms. However, {\color{blue}\S\ref{subsec:Laws in terms of Organizations}}  shows that these aspects differ across regulations, potentially causing problems for organizations.


\medskip
{\bf \noindent \ding{228} RQ3: How do differences in regulatory oversight and penalty structures result in risks to be faced by developers operating across multiple jurisdictions?} 

Data protection laws not only offer different things to the user and the organization collecting user's data, but {\color{blue}\S\ref{subsec:Laws in terms of Government}}  discusses that data protection laws have significantly different actions that the government/regulator takes when data protection laws are followed and/or violated by organizations.

{\color{black}

}

\bgroup
\def\arraystretch{1.1}
\begin{table}[t!]
  \scriptsize
\centering
\begin{tabular}{|p{3cm}|p{3cm}|p{1cm}|}
\hline
\textbf{Paper} & \textbf{Regulations considered} & \textbf{Year}\\
  \hline
  \hline


Zimmeck et al.~\cite{zimmeck_automated_2017} & COPPA; CalCOPPA (US) & 2017\\
\hline

Reyes et al.~\cite{reyes_wont_2018} & COPPA (US) & 2018\\
\hline

Zimmeck et al.~\cite{sebastian_zimmeck_maps_2019} & GDPR; COPPA & 2019\\\hline 

Jia et al.~\cite{jia_who_2019} & GDPR & 2019\\\hline

Mohan et al.~\cite{mohan_analyzing_2019} & GDPR & 2019\\
\hline

Ayalon et al.~\cite{10.5555/3361476.3361480} & GDPR & 2019\\
\hline

Fan et al.~\cite{fan_empirical_2020} & GDPR  & 2020\\
\hline

Singh et al.~\cite{singh_technical_2020} & PDPB (India) & 2020\\
\hline
Zhang-Kennedy et al.~\cite{zhang-kennedy_whether_nodate} & PIPEDA (Canada) & 2021 \\
\hline
Qamar et al.~\cite{qamar_detecting_2021} & GDPR; PDPA (Singapore) & 2021 \\
\hline

Nguyen et al.~\cite{nguyen_share_nodate} & GDPR & 2021 \\
\hline

Kollnig et al.~\cite{kollnig_fait_nodate} & GDPR; COPPA & 2021 \\
\hline

Nguyen et al.~\cite{nguyen_freely_2022} & GDPR & 2022\\
\hline

Koch et al.~\cite{koch_ok_nodate} & GDPR & 2023\\
\hline

Xiang et al.~\cite{xiang_policychecker_2023} & GDPR & 2023\\
\hline

Sassetti et al.~\cite{sassetti_assurance_2023} & GDPR & 2023\\
\hline

Ferreira et al.~\cite{ferreira_rulekeeper_2023} & GDPR & 2023\\
\hline

Herwanto et al.~\cite{herwanto_leveraging_2024} & GDPR & 2024\\
\hline


Krämer et al.~\cite{kramer_death_2024} & GDPR & 2024\\
\hline

Li et al.~\cite{li_are_2024} & GDPR & 2024\\
\hline

Liu et al.~\cite{liu_opted_2024} & GDPR; CCPA (US) & 2024\\
\hline

\end{tabular}
\caption{Overview of privacy compliance studies.} 
\label{tab:privacy_studies}
\BBB\BBB\BBB
\end{table}

\B
\subsection{Limitations of Using Localized Laws}
\label{subsec:Limitations of Using Localized Laws}
\BB



\noindent
\textbf{The majority of existing work focuses on GDPR.}  
A common theme among technical studies analyzing the privacy of digital systems is reliance on privacy regulations as a basis for threat modeling. 
As GDPR is extremely comprehensive, many studies refer to it as the gold standard, which explains why the majority of studies, as shown in {\color{blue}Table ~\ref{tab:privacy_studies}}, used GDPR as their legal framework baseline.  
However, this introduces limitations when privacy models are applied outside the conceptual and jurisdictional reach of such regulations, as the \emph{requirements change dramatically}, as we described in this work. 
We also note that this problem is not specific to GDPR---any study relying on a single regulation risks being overly specific. 
In a globalized world, such a practice introduces blind spots that limit the ability to generalize across regions and thus the applicability of research outcomes.

Below, we investigate how \emph{\textbf{differences in privacy regulations affect two significant limitations: variability in definitions of personal data and in consent/privacy requirements}}, which have been considered in most existing privacy analysis work.



\subsubsection{\bf Definitions of Personal and Sensitive Data}
\label{subsubsec:Definitions of Personal and Sensitive Data}

While data protection laws focus on personally identifiable information (PII), definitions of personal and sensitive data vary across regulations. 
Most regulations recognize names, addresses, and identification numbers as personal data; however, certain types of indirect identifiers, such as device metadata, behavioral data, or IP addresses, are treated differently across legal frameworks.

For example, GDPR (Article 4) defines PII as ``any information relating to an identified or identifiable natural person.'' 
In contrast, India's DPDPA states (Article 2(t)) that 
\emph{personal data means ``any data'' about an individual who is identifiable by or
in relation to such data}.
The phrase ``any data'' is ambiguous and leaves room for interpretation --- does it extend to handwritten records, video surveillance data, or AI-inferred attributes? 
Likewise, CCPA also includes ``household'' data, which is not covered by other regulations (Article 1798.140.(b)).

This inconsistency creates technical challenges when determining which categories of PII are in scope for privacy analysis --- a data protection approach that aligns with GDPR may still fall short of Indian privacy law.

Despite this issue, a substantial portion of reviewed studies operationalized personal/sensitive data using  GDPR's definition when detecting data leakage, classifying the lawfulness of data flows, or assessing privacy policy completeness ~\cite{fan_empirical_2020, jia_who_2019, mohan_analyzing_2019, ferreira_rulekeeper_2023, nguyen_freely_2022, nguyen_share_nodate}. 
As discussed, using one regulation (\textit{e}.\textit{g}., GDPR) for this definition lacks generalizability across the globe. For example, Nguyen et al. ~\cite{nguyen_freely_2022, nguyen_share_nodate} used GDPR’s definition of personal data. Applying this from the perspective of India's DPDPA and China's PIPL would not be able to include enough PII factors. 

Overall, relying on one or two regulations' definition of personal and sensitive data focuses on the risks present or considered in that jurisdiction, while omitting risks from other regulations. As a result, studies that focus on one local regulation (\textit{e}.\textit{g}., GDPR) might omit entire classes of data protected by different regulations. 

\subsubsection{\bf Consent and Privacy Expectations}
\label{subsec:consent_privacy_req}
Many studies have analyzed the notion of \textbf{\textit{consent}} and related requirements through the lens of: privacy policy completeness, run-time consent and privacy notices, and the alignment of collected data to natural language ~\cite{nguyen_freely_2022, li_are_2024, liu_opted_2024, koch_ok_nodate, kollnig_fait_nodate, xiang_policychecker_2023}. The majority of these studies grounded their threat models using GDPR's opt-in requirements, which requires consent to be freely given, informed, and unambiguous. 

As we will discuss in detail soon, consent models differ drastically across regions. 
For example, California's CCPA focuses on opt-out rights, Brazil's LGPD 
permit processing based on legitimate interest, and India's DPDPA introduces special provisions for persons with disabilities. 
This shows that using a single regulation (\textit{e}.\textit{g}., GDPR) definition of lawful consent as a global standard may misclassify lawful practices and actions, or even miss compliance with different regulations (\textit{e}.\textit{g}., California's CCPA opt-out rights vs others' opt-in rules). 

A similar issue arises when characterizing the notion of \textbf{\textit{privacy breach}}. Data protection laws focus on restricting access to personal data but do not always specify which technical actions or system behaviors may result in such a breach. This ambiguity complicates the assessment of data collection appropriateness.

For instance, a health and fitness application that measures walking distance may access a smartphone's accelerometer and GPS to track movement. However, if the same app also requests access to the microphone or ambient light sensor, it may commit a privacy violations under certain regulations. Under Article 5(1)(c) of GDPR, data collection should be ``limited to what is necessary in relation to the purposes for which they are processed'' (data minimization principle). 
If an app accesses unnecessary sensors without a justified purpose, it could violate GDPR. 
In contrast, Brazil's POPIA (Article 9) requires ``personal information must be processed 
lawfully and in a reasonable manner that does not infringe the privacy of the data subject.'' This leaves room open to collect other sensor data via the health app, unless mixed Article 10, which states ``personal information may only be processed if, given the purpose for which it is processed, it is adequate, relevant and not excessive.''

Any technique attempting to measure breaches in the wild may therefore produce different outcomes depending on the relevant legislation.

\bgroup
\def\arraystretch{1.25}
\begin{table*}[t]
\scriptsize
\centering
\begin{tabular}{|l|p{4.4cm}|p{5.7cm}|p{3cm}|}
\hline
\textbf{Data Life-Cycle Phase} 
& \textbf{Users ({\color{blue}\S\ref{subsec:Laws in terms of the Users})}} 
& \textbf{Organizations} ({\color{blue}\S\ref{subsec:Laws in terms of Government})}
& \textbf{Government} ({\color{blue}\S\ref{subsec:Laws in terms of Government})}\\ \hline

Collection 
& Right to consent and right to revoke consent 
& 
Rules for purpose limitation, data minimization, and transparency ({\color{blue}\S\ref{subsubsec:Core Processing Principles}})
& \multirow{6}{*}{\parbox{3cm}{\centering
Investigate and audit the records of processing, and different types of penalties
}} \\ \cline{1-3}

Storage 
& Right to access and right to data portability 
& Rules for security, accuracy, and data retention ({\color{blue}\S\ref{subsubsec:Core Processing Principles}}) 
& \\ \cline{1-3}

Processing 
& Right to object and right to restriction of processing 
& Rules for accountability ({\color{blue}\S\ref{subsubsec:Accountability}})
& \\ \cline{1-3}

Sharing 
& Right to restriction of processing 
& Rules for cross-jurisdiction data transfer ({\color{blue}\S\ref{subsubsec:Cross Jurisdiction Data Transfer}})
& \\ \cline{1-3}

Deletion 
& Right to delete 
& Rules for deletion 
({\color{blue}\S\ref{subsubsec:Data Deletion}})
& \\ \cline{1-3}

Data Bequeath 
& Right to nominate 
& Rules for accountability  ({\color{blue}\S\ref{subsubsec:Accountability}})
& \\ \hline

\end{tabular}
\caption{Mapping of data life cycle phases to individual/user, organization, and government roles.}
\label{tab:data_lifecycle_mapping}
\BBB\BBB\BBB
\end{table*}

\B
\subsection{Choice of Regulations in the Paper}
\label{subsec:Choice of Regulations in the Paper}
\BB

In selecting privacy regulations, we address the tension between keeping the legal analysis workload manageable and understandable, while ensuring sufficient representativeness of worldwide laws. 
Our selection process is as follows. First, we add GDPR to the set because of its widespread use in relevant research. Then, we select additional regulations subject to the constraint that each new regulation covers a continent not yet in our set.\footnote{\scriptsize We consider China separately from the rest of Asia, due to its cultural specificities and outsized impact on worldwide commerce.} When adding each new regulation, we aim for populous countries with established data protection laws. We stop when we achieve approximately 40\% coverage of the world population. 
{\color{black}The population criterion was to ensure broad data subject (or called user) coverage while maintaining analytical depth. Continental consideration was introduced to encourage regulatory diversity.  We selected China and India because of the scale of the data subjects and the distinctiveness of their regulatory approaches. However, ultimately achieving our goal is not dependent on any specific choice of regulations, as long as the choice showcases regulatory diversity.}

The resulting set includes: GDPR (EU), CCPA (California, US), DPDPA (India), PIPL (China), POPIA (South Africa), and LGPD (Brazil), covering $\approx$38.3\% of the world population. 
While we acknowledge that this set does not include other representative laws, we argue that it suffices to qualitatively highlight the limitations stemming from only considering individual regulations.

}

\B
\subsection{Summary 
of the Paper}
\BB

This paper examines how modern privacy regulations govern personal data and analyzes the challenges these regulations pose for organizations, individuals, and regulators.
We adopt a data life-cycle perspective to structure this analysis for two reasons.
First, these data protection laws primarily regulate automated or systematized personal data processing, which naturally unfolds across distinct phases, such as data collection, storage, processing, sharing, deletion, and bequeathal.
Second, regulatory obligations and user rights are not meaningful in isolation; rather, they are triggered, constrained, and enforced depending on when and how personal data is handled within the data life cycle.
In practice, organizations operationalize privacy through technical mechanisms deployed across these phases, while individuals exercise their rights in response to phase-specific data handling activities.

With this perspective, {\color{blue}\S\ref{sec:related}} discusses related work and
{\color{blue}\S\ref{sec:Privacy Laws Data Life Cycle and User Rights}} systematically maps major privacy regulations to the data life-cycle phases. 
Building on this mapping, 
{\color{blue}\S\ref{subsec:Laws in terms of the Users}} analyzes implications for individuals, highlighting variations across data protection laws in the defining personal data, sensitive data, children's data, user rights, and consent models.
Next, {\color{blue}\S\ref{subsec:Laws in terms of Organizations}}  examines implications for organizations by identifying the essential functional requirements that organizations must implement at each phase of the data life cycle to achieve compliance.
Finally, {\color{blue}\S\ref{subsec:Laws in terms of Government}}  compares government authority roles across jurisdictions, focusing on supervisory power differentials, and mechanisms of investigation and enforcement.

\bgroup
\def\arraystretch{1.25}
\begin{table*}[h]
    \centering
   \scriptsize  
   \begin{tabular}{|p{0.5cm}|p{0.7cm}|p{1.6cm}|p{5.2cm}|p{2.2cm}|p{5.4cm}|}\hline

        \textbf{Laws} & \textbf{Article/ Section} & \textbf{Applicable countries/ regions} & \textbf{Organization presence in jurisdictional region} & \textbf{Individual residence in jurisdictional region} & \textbf{Not applicable to} \\\hline

        GDPR & 3 & EU & Applies if the organization processes data of EU residents, even if located outside the EU & Applies to individuals residing in the EU & Data collected for personal or household activities, or by authorities for crime prevention, investigation, prosecution, or public security \\\hline

        CCPA & 1798.140 & California (U.S.) & Applies to businesses processing data of California residents, whether or not based in California, and having annual gross revenues exceeding \$25M or deriving 50\%+ revenue from selling/sharing personal data & Applies to California residents & Not explicitly mentioned \\\hline

        DPDPA & 3 & India & Applies to businesses processing data of Indian residents, regardless of businesses' location & Applies to Indian residents & Data collected by an individual for personal or domestic purposes, or personal data made public by the individual, Art.~3(c) \\\hline

        PIPL & 3 & China & Applies to organizations processing data of individuals residing in China, regardless of organizations' location & Applies to Chinese residents & Not explicitly mentioned \\\hline

        LGPD & 3 & Brazil & Applies to businesses processing data of Brazilian residents, regardless of businesses' location & Applies to Brazilian residents &  Data collected  by a person for private, non-economic purposes; or for journalistic, artistic, public safety, national defense, state security, or investigation purposes, Art.~4 \\\hline

        POPIA & 3 & South Africa & Applies to organizations processing data of individuals domiciled in South Africa, regardless of organization's location & Applies to South African residents & Not explicitly mentioned \\\hline

    \end{tabular}
    \caption{Applicability of Data Protection Across Jurisdictions}
    \label{tab:Applicability of Privacy Laws Across Jurisdictions}
    \BBB
\end{table*}
\egroup

\bgroup
\def\arraystretch{1.25}
\begin{table*}[!h]
\BBB
    \centering
    \scriptsize
    \begin{tabular}{|p{0.5cm}|p{4cm}|p{5.4cm}|p{6cm}|}
        \hline
        \textbf{Law} & \textbf{Personal Data} & \textbf{Examples of Personal Data} & \textbf{Sensitive Data} \\\hline

        {GDPR} & Any information relating to an identified or identifiable natural person, Art.~4 & 
        Name, identification number, location data, online identifier, or factors specific to physical, physiological, genetic, mental, economic, cultural, or social identity & Racial or ethnic origin, political opinions, religious or philosophical beliefs, trade union membership, genetic data, biometric data, Art.~9 \\\hline

        {CCPA} & Information that identifies, relates to, describes, is reasonably capable of being associated with, or could reasonably be linked, directly or indirectly, with a particular consumer or household, Art.~1798.140 
        & Real name, alias, unique personal identifier, online identifier, IP address, internet activity, geolocation data, account name, employment information, education data, biometric data (\textit{e}.\textit{g}., DNA, imagery of the iris, retina, fingerprint, face, hand, palm, vein patterns, and voice recordings, faceprint) 
        & Social security number, driver's license, state ID, passport number, account login, financial account details, precise geolocation, racial or ethnic origin, citizenship or immigration status, religious or philosophical beliefs, union membership, genetic data, neural data, Art.~140(ae) \\\hline

        {DPDPA} & Any data about an individual who is identifiable by or in relation to such data,  Art.~2(t) & Not specified & Not specified$^\dag$ \\\hline

        {PIPL} & Various information related to an identified or identifiable natural person recorded electronically or by other means, Art.~4 & Not specified & Biometrics, religious belief, specific identity, medical health status, financial accounts, and the person's whereabouts, Art.~28 \\\hline

        {LGPD} & Information regarding an identified or identifiable natural person (Art.~5I) & Not specified & Racial or ethnic origin, religious belief, political opinion, trade union or religious, philosophical, or political organization membership, health or sex life data, genetic or biometric data, Art.~5II \\\hline

        {POPIA} & Information relating to an identifiable, living, natural person, Art.~1 & Gender, marital status, education information, identifying number, symbol, email address, physical address, phone number, location data, online identifier, person's name & 
        Religious or philosophical beliefs, race or ethnic origin, trade union membership, political persuasion, health or sex life, biometric information, criminal behavior, Art.~26 \\\hline

    \end{tabular}
    \caption{Definitions of personal and sensitive data across data protection laws. Note: $\dag$: DPDPA is the only law that focuses on persons with disability, see Art.~2(j).}
    \label{tab:data_definitions}
    \BBB\BBB
\end{table*}
\egroup

\section{Related Work}
\label{sec:related}
\BB
{\color{black}Several privacy frameworks, such as Contextual Integrity~\cite{nissenbaum2004privacy,DBLP:conf/sp/BarthDMN06} and Interdependent Privacy~\cite{IDP,InterdependentPrivacy}, provide formal approaches for defining and reasoning about privacy. 
Contextual Integrity conceptualizes privacy as the appropriate flow of information within specific social contexts, governed by norms that specify actors, data subjects, senders, recipients, information types, and transmission principles. 
Interdependent Privacy emphasizes that an individual's privacy is not solely determined by their own actions but is also influenced by others' disclosures and behaviors.

While these frameworks offer valuable conceptual foundations, they do not ensure enforceable protection in practice, as they do not specify enforcement mechanisms, legal obligations, or remedies. 
In contrast, the data protection laws considered in this paper operationalize privacy through enforceable user rights, obligations imposed on data controllers and collectors, and mechanisms for regulatory oversight.

Commercial platforms, e.g., 
DLA Piper~\cite{DLA} and OneTrust DataGuidance~\cite{onetrust}, provide jurisdiction-based summaries. However, they are also limited, e.g., by not discussing consent and user rights, and by functioning only in operational categories to realize rights.  Overall, the platforms do not provide straightforward ways for researchers to understand the differences among laws.
 }

\section{Laws and Data Life Cycle}
\label{sec:Privacy Laws Data Life Cycle and User Rights}
\BB
A systematic understanding of data protection laws starts with the data life cycle, which captures the sequential stages through which personal data passes. 
At each stage of the data life cycle, data protection laws follow an \emph{actor-centric} model, \textit{i}.\textit{e}., they place constraints/ rules/ rights on the three main actors --- the user/ individual/ personal whose data is collected, the organization that collects the individual's data, and the government that enforces the data protection law. 
In particular, the data protection laws provide:
(\textit{i})~the rights for individuals whose data is being processed to manage their data, 
(\textit{ii})~the rules for organizations to handle individuals' data, and 
(\textit{iii})~the powers to the governments and regulatory authorities to supervise and enforce compliance.
In short, these laws impose constraints/rules to ensure that individuals' data is collected, processed, stored, shared, and deleted in a responsible and legally compliant manner.

{\color{blue}Table~\ref{tab:data_lifecycle_mapping}} provides a mapping of each life-cycle phase to (\textit{i})~the rights granted to users, (\textit{ii})~the rules imposed on organizations, and (\textit{iii})~the enforcement responsibilities for the governments. 
This table shows how user rights and organizational duties are distributed across different stages of the data life cycle.

Below, we outline the stages of the data life cycle, and then, the remaining paper will explain each cell of the table in detail.

\begin{enumerate}[leftmargin=0.01in] 
\item 
\noindent
\textbf{Data collection (Row~1).} Personal data is collected from individuals through various methods such as online forms, surveys, or interactions with services and applications. 
Data protection laws ensure that individuals retain control over their data from the moment it is collected. 
Rights related to this phase include the right to consent and the right to revoke consent, while organizations are required to follow rules of purpose limitation, data minimization, and transparency.

\item 
\noindent
\textbf{Data storage (Row 2).} Once collected, data is stored for varying periods based on legal requirements or organization/business needs. Data protection laws regulate how data must be stored, protected, and managed to prevent unauthorized access. 
Rights related to this phase include the right to access and the right to data portability, 
while organizations are required to follow security, accuracy, and data retention rules.

\item 
\noindent
\textbf{Data processing (Row~3).} Data is processed for various purposes, such as analysis, reporting, or decision-making. During this stage, individuals retain specific rights—including the right to object and the right to restriction of processing—to ensure their data is handled in accordance with their consent, while organizations are required to follow accountability rules.

\item 
\noindent
\textbf{Data sharing (Row~4).} At this stage, personal data may be shared with third parties, including service providers and business partners. Data protection laws regulate how data should be shared and ensure individuals have control over how their information is distributed. 
Rights related to this phase include the right to restriction of processing, 
while organizations are required to follow cross-jurisdiction data transfer rules.

\item 
\noindent
\textbf{Data deletion (Row~5).} When personal data is no longer needed for its original purpose, it should be deleted/anonymized. 
Data protection laws empower individuals to control when and how their data is erased, ensuring it is not retained longer than necessary.
Rights related to this phase include the right to erasure/deletion, 
while organizations are required to follow deletion rules.

\item 
\noindent
\textbf{Data bequeath (Row~6).}
Data protection laws extend data governance beyond an individual's lifetime by allowing posthumous or delegated control over personal data. This phase governs the nomination of a representative who may exercise data-related rights on behalf of an individual in the event of death or incapacity. 
Rights related to this phase include the right to nominate, while organizations are required to adhere to accountability rules.
\end{enumerate}

\section{Diversity in Data Protection Laws}
\label{sec:Diversity in Privacy Laws}
\BB

Data Protection laws pose significant compliance challenges, particularly for organizations operating across multiple jurisdictions, as well as individuals traveling from one jurisdiction to another. 
No universal data protection law exists; instead, each country or region establishes its own regulations, which differ based on factors such as jurisdictional boundaries, an individual's physical presence, an organization's presence, and its turnover. {\color{blue}Table~\ref{tab:Applicability of Privacy Laws Across Jurisdictions}} compares these laws in terms of their jurisdictional applicability. 
For example, GDPR applies to any EU residents, even if they are currently outside the EU, whereas CCPA introduces business-specific thresholds. Laws such as the PIPL, DDPA, LGPD, and POPIA apply to subjects who are resident or domiciled within their borders. 

Beyond jurisdiction, data protection laws also vary substantially in definitions of data, consent requirements, user rights, cross-border data transfer rules, penalties for violations, and mechanisms for enforcing compliance (and due to which leads to improper implementation and/or violation of the laws~\cite{fan_empirical_2020, li_are_2024, mohan_analyzing_2019}).

To systematically study these differences, in this section, we classify data protection laws according to three main actors: 
the \emph{user/individual} whose data is collected ({\color{blue}\S\ref{subsec:Laws in terms of the Users}}), 
the \emph{organization} that collects and processes this data ({\color{blue}\S\ref{subsec:Laws in terms of Organizations}}), and 
the \emph{government} that defines privacy regulations and enforces compliance ({\color{blue}\S\ref{subsec:Laws in terms of Government}}).

\emph{\bf Note that the laws we compare in this paper are divided into articles and/or sections. For simplicity, we use the word `article' in this paper.}

\bgroup
\def\arraystretch{1.25}
\begin{table*}[!h]
    \centering
    \scriptsize
    \begin{tabular}{|l|p{1.5cm}|p{3cm}|p{7cm}|p{2.2cm}|}\hline
        
        \textbf{Law} & \textbf{Article} & \textbf{Age threshold for consent} & \textbf{Consent mechanism} & \textbf{Verification of consent} \\\hline
        
        {GDPR} & Art.~8 & Under 16 years (can be lowered to 13 by member states) & Parental consent & Yes \\\hline
        
        {CCPA} &  Art.~1798.120 and 999.330 & 
        Under 13 years & Custodial parent or guardian required for children under 13; and for ages 13-16, the child can provide consent & Yes \\\hline
        
        {DPDPA} & Art.~9 & Under 18 years & Consent from a parent or lawful guardian & Yes \\\hline
        
        {PIPL} &  Art.~28, 31 & Under 14 years & Consent of parents or other guardians & No \\\hline
        
        {LGPD} &  Art.~14 & Under 12 years & Consent of one parent or legal representative & Yes \\\hline
     
        {POPIA} & Art.~35 & Under 18 years & Consent of a competent person & No \\\hline
        
    \end{tabular}
    \caption{Children's data protection requirements across data protection laws.}
    \label{tab:child_data_protection}
    \BBB\BBB\BB
\end{table*}\egroup

\B
\subsection{Laws in terms of the Users}
\label{subsec:Laws in terms of the Users}
\BB

\textbf{Objectives.} 
Analysis of data protection laws must begin with the individual, as the user is the primary subject whose data triggers the entire data life cycle. Thus, the objective of this section is to first examine how data protection laws categorize user data as personal, sensitive, or children's data in~{\color{blue}\S\ref{subsec:Personal Data, Sensitive Data, and Child Data}}. 
Then, our objective is to discuss how such data classifications specify the rights granted to individuals and the mechanisms for controlling their data, including its collection, use, and deletion, which we will discuss in subsequent subsections~{\color{blue}\S\ref{subsubsec:User Rights in the Data Life Cycle: Purpose and Significance}} and~{\color{blue}\S\ref{subsubsec:Consent Models}}.

\bgroup
\def\arraystretch{1.25}
\begin{table*}[!h]
    \centering
    \scriptsize
    \renewcommand{\arraystretch}{1}
    \begin{tabular}{|p{5.2cm}|p{0.8cm}|p{4.3cm}|p{0.8cm}|p{0.9cm}|p{0.7cm}|p{0.7cm}|}
      \hline
      
        \textbf{Privacy Right} & \textbf{GDPR} & \textbf{CCPA } & \textbf{DPDPA} & \textbf{PIPL } & \textbf{LGPD } & \textbf{POPIA} \\\hline

     
        Right to Consent & Art.~6, 13 & Default opt-in. Opt-out is available, Art.~1798.120 & Art.~5, 6 & Art.~13, 14 & Art.~7 & Art.~11 \\\hline
        
        Right to Revocation of Consent & Art.~7(3) & No & Art.~6(4) & Art.~15 & Art.~8(5) & Art.~11(3) \\\hline
        
        Right to Access / Right to Explanation of Handling Policies & Art.~15 & Art.~1798.110, Art.~115 & Art.~11 & Art.~45 & Art.~18 & Art.~23 \\\hline
        
        Right to Data Portability & Art.~20 & No & No & Art.~45 & Art.~18 & No \\\hline
        
        Right to Erasure/Deletion (Right to Be Forgotten) & Art.~17 & Art.~1798.105 & 12(1) & Art.~47 & Art.~18 & Art.~24 \\\hline
        
        Right to Rectification & Art.~16 & Art.~1798.106 & No & Art.~12(1) & Art.~46 & Art.~24 \\\hline
        
        Right to Object & Art.~21 & Art.~1798.120, 121 & Art.~13$^\ddag$ & Art.~44 & Art.~18 & Art.~11(3) \\\hline
        
        Right to Restrict Processing & Art.~18 & Art.~1798.120 & No & Art.~44 & No & Art.~11(3) \\\hline
        
        Right to Nominate & No & No & Art.~14 & Art.~49$^\dag$  & No & No \\\hline

        Right to Non-Discrimination & No & Art.~1798.125 &No & No & No & No \\\hline
        
        \end{tabular}
    \caption{Comparison of privacy rights. {$\ddag$: Under ``Grievance redressal'' rights. $\dag$: for deceased individuals.}}
    \label{tab:privacy_rights}
    \BBB\BBB
\end{table*}\egroup

\subsubsection{\bf Personal Data, Sensitive Data, and Child Data}
\label{subsec:Personal Data, Sensitive Data, and Child Data}

Data protection laws classify users' data into three categories: personal, sensitive, and children's data, which are differentiated based on the specificity of data and the age of the user, as we discuss below:

\medskip
\noindent 
\textbf{Personal data} refers to any information that directly or indirectly identifies an individual. While most data protection laws share a common foundation in terms of personal/user/individual data, their definitions vary in scope. Some laws adopt broader definitions; others provide only vague definitions. 
For instance, GDPR and South Africa's POPIA adopt broader definitions, explicitly including indirect identifiers such as online identifiers (\textit{e}.\textit{g}., IP addresses and cookies), while India's DPDPA and China's PIPL do not clearly specify the meaning of personal data.

\medskip
\noindent 
\textbf{Sensitive personal data} provides fine-grained information about very specific data, whose collection may impact an individual's dignity, safety, and privacy.  Examples of sensitive data typically include information on racial or ethnic origin, political opinions, religious or philosophical beliefs, trade union membership, genetic and biometric data, and financial details. Only India's DPDPA does not deal with such sensitive data; nevertheless, DPDPA deals with  \emph{\textbf{persons with disabilities}} and allows data collection from them if a lawful guardian provides verifiable consent.


\medskip
\noindent 
\textbf{Children's data} is defined in terms of the age of minors, which varies according to different laws from 13 to 18. The laws also enforce that whenever data from a minor is collected, a  consent must be obtained from the lawful guardian, not from the child, and there should be a mechanism to verify that the consent is really obtained from lawful guardians.

{\color{blue}Tables~\ref{tab:data_definitions}} 
compares different laws in terms of the definition of personal and sensitive data, along with explicit examples of personal data. {\color{blue}Table~\ref{tab:child_data_protection}} compares different laws in terms of children's data, the way of finding children, and their consent model.  

\smallskip
\textbf{Note.} While all the laws strictly define personal data and mechanisms to handle data, only India's DPDPA (Art.~14(1)) and China's PIPL (Art.~49) also provide rights to the individual to nominate someone on their behalf in case of their death or incapacity.

\bgroup
\def\arraystretch{1.25}
\begin{table*}[!h]
\BBB
    \centering
    \scriptsize
    \begin{tabular}{|p{0.5cm}|p{0.7cm}|p{2.4cm}|p{12.5cm}|}
        \hline
        \textbf{Law} & \textbf{Article} & \textbf{Consent Model} & \textbf{Key Requirements} \\
        \hline
        
        {GDPR}  & 6, 7 & Opt-in, Art.~6(1a), 7(1) & Freely given, specific, informed, and unambiguous indication of the individual's wishes, signified by a statement or clear affirmative action agreeing to the processing of personal data. \\
        \hline
        
        {CCPA} & 1798.120 & Opt-out & Businesses must provide a clear and conspicuous ``Do Not Sell My Personal Information'' link, allowing consumers to opt-out of data sales. Parental consent is required for selling data of children under 13, and affirmative consent is required for those aged 13-16. \\
        \hline
        
        {DPDPA}  &6 & Opt-in, Art.~6(1) & Free, specific, informed, unconditional, and unambiguous consent, given through a clear affirmative action. Consent must be limited to necessary personal data for a specified purpose.$\dag$ \\
        \hline
        
        {PIPL}  & 7,~14,~17 & Opt-in,  Art.~(14) & Requires openness and transparency. Organizations must disclose processing rules and provide information in a clear and understandable manner. Consent must be voluntary, explicit, and based on full knowledge. \\
        \hline
        
        {LGPD}  & 5-10 & Opt-in, Art.~5(9) & Requires free, informed, and unambiguous consent. Allows processing based on legitimate interest in certain cases without consent, but explicit consent is required for sensitive data. \\
        \hline
        
        {POPIA}  & 11 & Opt-in (for sensitive data) and opt-out (for marketing) & Requires explicit consent for processing sensitive personal information. Allows opt-out mechanisms for direct marketing. Individuals must be informed about the purpose of data collection. \\
        \hline
        
    \end{tabular}
    \caption{Comparison of consent models across data protection laws. {Note: $\dag$:  DPDPA has a special provision for persons with disabilities, and only the lawful guardian should provide verifiable consent for persons with disabilities, Art.~9(1).}}
    \label{tab:consent_models}
    \BBB\BBB
\end{table*}\egroup

\bgroup
\def\arraystretch{1.25}
\begin{table*}[!h]
\scriptsize
    \centering
    \begin{tabular}{|p{0.7cm}|l|l|l|p{7cm}|p{4cm}|p{5cm}|p{8cm}|}
        \hline
        \textbf{Law} & \textbf{Article} & \textbf{Consent Type} & \textbf{Time-bound} 
        \\
        \hline

        {GDPR} & Art.~4(11), 6(1a), 7(1) &  Specific Consent, Art.~7(1) & 
        Yes, Art.~7(3)
        \\\hline

{CCPA} & Art.~1798.120 & Broad General Consent &No\\
        \hline

  {DPDPA} & Art.~6(1)& Specific Consent, Art.~6(1)& No (but the right to withdraw her consent at any time), Art.~6(4),6(7) \\\hline
            
        {PIPL} & Art.~5, 13, 15 & Specific Consent &
       No, (but withdraw his/her consent), Art.~5
        \\\hline

        {LGPD} & Art.~8, 9 & Moderately Specific & Yes, Art.~9(2), and consent may be revoked at any time, Art.~8(5) 
        \\\hline

        {POPIA} & Art.~13, 14 & Context-Dependent Consent &Yes, Art.~14
        \\\hline

      \end{tabular}
    \caption{Granularity of consent. }
    \label{tab:consent_granularity}
    \BBB\BBB
\end{table*}\egroup

\bgroup
\def\arraystretch{1.25}
\begin{table*}[!h]
\scriptsize
    \centering
    \begin{tabular}{|l|l|p{4.5cm}|p{3.2cm}|l|p{2.3cm}|p{6cm}|l|}
        \hline
        \textbf{Law} & \textbf{Consent Withdrawal} &  \textbf{Key Requirements for Withdrawal} & \textbf{Duration for Data Deletion After Consent Withdrawal} & \textbf{Consent expiration}\\
        \hline
        
        {GDPR}  & Right to Withdraw Consent at Any Time, Art.~7(3) & 
        Easy to withdraw as to give consent & No, Art.~17(1b) & Yes, Art.~5(1b), 25(2)
        
        \\
        \hline

        {CCPA}  & Right to opt out of sale or sharing, Art.~1798.120 & N/A & 45 days, Art.~1798.130(a)(2) & No\\
        \hline

  {DPDPA}  & Right to Withdraw Consent at any time, Art.~6(4) &  Ease of doing so being comparable to the ease with which such consent was given  & No, Art.~6(6), 8(7) & No 
  \\\hline
        
        {PIPL}  & Right to Withdraw Consent, Art.~15 & Convenient means to withdraw consent  & No 
        & No

      \\\hline

        {LGPD}  & Consent may be revoked, Art.~8(5) & 

         Free of charge procedure & No
         & Yes, Art.~9(2)\\
        \hline
        
        {POPIA}  & Withdraw consent, Art.~11(2b)  &

       Not given & No 
       & Yes, Art.~13(1), 14(1)\\
        \hline

    \end{tabular}
    \caption{Consent withdrawal rules across  laws.}
\label{tab:consent_withdrawal_expiration}
\BBB\BBB\BB
\end{table*}\egroup

\subsubsection{\bf User Rights}
\label{subsubsec:User Rights in the Data Life Cycle: Purpose and Significance}

Data protection laws offer individuals several rights that empower them to control their personal data throughout its lifecycle, which includes collection, storage, processing, sharing, and deletion. 
These rights ensure that data collection and processing are conducted transparently, with user consent, and that individuals can access, modify, or delete their data as needed. 
By enforcing these rights, privacy regulations offer benefits such as transparency, accountability, prevention of misuse, and enhanced trust in data management practices. 
{\color{blue}Table~\ref{tab:privacy_rights}} compares ten user rights across different laws. No law grants all 10 such rights to the user. Only China's PIPL offers 9 out of 10, while GDPR and India's DPDPA offer 8 out of 10.

Below, we discuss these rights and their benefits to the user.

\medskip
\noindent 
\textbf{Right to consent:} ensures transparency by offering individuals the ability to understand and control how their personal data is collected and processed. 
Consent must be given through a clear, affirmative act that is freely given, specific, informed, and unambiguous. This can be expressed through a written statement, electronic means, or an oral declaration. 
Without valid consent, data processing is generally prohibited unless justified by legitimate interests or other legally recognized grounds.

\medskip
\noindent
\textbf{Right to revocation of consent:} ensures flexibility to individuals by allowing them to revoke/withdraw their consent at any time. Organizations should implement easy-to-use mechanisms, \textit{e}.\textit{g}., opt-out links. Once consent is revoked, data processing must cease unless another valid legal basis justifies its continuation.

\medskip
\noindent
\textbf{Right to access:} ensures transparency and accountability by allowing individuals to request and receive a copy of their personal data that is held by an organization, along with details about how their data is being used. Organizations must provide individuals with a copy of their data upon request within a reasonable timeframe, along with details on how and why it is being processed. PIPL considers the same right under ``Right to explanation of handling policies.''

\medskip
\noindent
\textbf{Right to rectification:} ensures accuracy of personal data by offering individuals the right to request corrections or updates to any inaccurate or incomplete personal data. Organizations should provide individuals with accessible means to promptly correct and update their personal data.

\medskip
\noindent
\textbf{Right to deletion:} ensures control over the personal data by offering individuals the right to request the deletion of their personal data under certain conditions, such as when data is no longer necessary for its original purpose or when consent is revoked. Organizations should implement procedures to delete personal data upon user request when it is no longer necessary for processing.

\medskip
\noindent
\textbf{Right to object:} ensures individuals' autonomy by allowing them to refuse their personal data processing, particularly for purposes such as marketing or profiling. Organizations must provide individuals with an easy way to opt-out of specific data processing activities.

\medskip
\noindent
\textbf{Right to restriction of processing:} enables individuals to request a temporary restriction in the processing of their data under certain circumstances, such as when the accuracy of the data is disputed, when the processing is unlawful and the individual opposes to its erasure, or when the data is needed for legal claims, even if the organization no longer requires it for its own purposes.

\medskip
\noindent
\textbf{Right to data portability:} ensures control over the personal data by offering individuals the right to receive their personal data in a structured, commonly used, and machine-readable format, and to transfer it to another data controller. This right facilitates more effortless data transfer between providers.

\medskip
\noindent
\textbf{Right to nominate:} enables individuals to designate another person to exercise their rights in the event of death or incapacity, such as mental unsoundness or physical infirmity. This right ensures that an individual's data protection rights can still be managed by a nominated representative if they are no longer able to do so themselves. This right is provided explicitly under DPDPA (India, Article~16) and PIPL (China, Article~49).

\medskip
\noindent
\textbf{Right to non-discrimination:} protects individuals from being denied goods, services, or benefits, or being charged different prices or offered different levels of service simply because they exercised their privacy rights. This ensures that individuals are not penalized or treated unfairly for asserting their data protection rights, such as opting-out of data collection or requesting data deletion. This right is included under CCPA.

\bgroup
\def\arraystretch{1.25}
\begin{table*}[ht]
\scriptsize
\centering
\begin{tabular}{|p{0.5cm}|p{1.85cm}|p{1.6cm}|p{1.7cm}|p{2.1cm}|p{1cm}|p{2cm}|p{1.4cm}|p{2cm}|l|}
\hline

\textbf{Law} & \textbf{Purpose Limitation} & \textbf{Fairness} &\textbf{Transparency} &\textbf{Data Minimization} & \textbf{Accuracy} & \textbf{Storage Limitation} & \textbf{Security} & \textbf{Accountability} \\ \hline

GDPR & Art.~5(1b) & Art.~5(1a)& Art.~5(1a), 12, 13, 14 &  Art.~5(1c), 25(1) &   Art.~5(1d) &   Art.~5(1e) &   Art.~5(1f), 25, 32 &   Art.~5(2), 30 \\ \hline

CCPA & Art.~1798.100(b),(c) & Art.~1798.125& Art.~1798.130&  N/A & N/A&N/A & N/A&N/A \\ \hline

DPDPA &  Art.~4, 7 & Not explicit & Art.~5 & Not explicit, but Art.~6(1) & 12 & Not explicit, but Art.~8(8) & Art.~8(5) & Not explicit, but Art.~8 \\ \hline

PIPL  &   6 &Not explicit (but good faith, Art.~5) & Art.~7 and 50 &   Art.~6 &   Art.~8, 46 &   Art.~19 &   51(3) & Not explicit, but   Art.~5 (lawfulness), 48, 54 \\ \hline

LGPD  &   Art.~6(1) & Art.~6(9) &   Art.~6(6) &   Art.~6(3) &   Art.~6(5) &   Art.~6(7) &   6(6), 46 &   Art.~6(10) \\ \hline

POPIA & Art.~13, 18 & Not explicit & Art.~17 (Openness) & Art.~10 &  Art.~16,24 & Not explicit, but Art.~14 &  Art.~19, 21 &  Art.~23 \\ \hline

\end{tabular}
\caption{Core processing principles across data protection laws.}
\label{tab:core-principles}
\BBB\BBB\BBB
\end{table*}\egroup

\subsubsection{\bf Consent Models}
\label{subsubsec:Consent Models}

Consent forms a fundamental legal basis for data processing in all data protection laws; however, consent's  definition, scope, and enforcement differ significantly. 
{\color{blue}Table~\ref{tab:consent_models}} compares different laws with respect to the consent model (opt-in vs. opt-out) and the consent requirements for providing information in the consent form, such as non-ambiguity and no charges for the consent form. 

Note that \emph{except for CCPA, all the laws require opt-in, \textit{i}.\textit{e}., explicitly providing consent for their data collection by organizations}, while CCPA instead requires explicit opt-out.

Furthermore, consent can be specific to a particular task or general to all tasks carried out by the organization using the collected data, as we discuss below.

\medskip
\noindent
\textbf{Granularity of consent (general vs. specific consent).}

Granularity of consent refers to the level of detail and specificity with which individuals are asked to grant permission for the processing of their personal data. 
\emph{\textbf{General consent}} is a broad, overarching agreement that allows an organization to collect and use personal data for multiple purposes without requiring the individual to approve each specific use. 
In contrast, \emph{\textbf{specific consent}} is narrowly defined and purpose-linked, requiring the individual to explicitly agree to each particular way their data will be processed. 
The purpose of differentiating between general and specific consent is to give users greater control and awareness over how their personal information is handled. 

{\color{blue}Table~\ref{tab:consent_granularity}} compares laws on the consent type and the duration for a given consent. Note that 
GDPR, India's DPDPA, and China's PIPL require specific consent.
GDPR, LGPD, and POPIA allow data processing for a limited time under a given consent, while others require users to exercise their right to withdraw consent to prevent the organization from collecting their data. 

\medskip
\noindent
\textbf{Consent withdrawal and expiration}
are safeguards to ensure individuals retain control over their personal data. 
While both mechanisms restrict the scope of consent, they operate differently. 
\emph{\textbf{Consent withdrawal}} empowers individuals to actively revoke their consent at any time, obligating organizations to cease further collection or use of data based on that consent. 
Importantly, many laws distinguish between stopping future collection and the handling of data already obtained, which may still be processed under other lawful bases. 
\emph{\textbf{Consent expiration}}, in contrast, is linked to the principle of purpose limitation: consent is valid only for the specific purpose and duration for which it was originally given, and it lapses automatically once that purpose has been fulfilled, without requiring any action from the individual. 

Data protection laws vary in how they regulate these mechanisms; for example, some explicitly require withdrawal to be as simple as giving consent, prohibit charging fees, or specify the timeframe within which processing must cease after withdrawal.
{\color{blue}
Table~\ref{tab:consent_withdrawal_expiration}} compares data protection laws in terms of consent withdrawal mechanisms and their requirements, duration for data deletion after consent revocation, and options for consent expiration.

\subsubsection{\bf Lessons}
\label{subsubsec:lesson user terms}
While data protection laws share a common goal of providing users with meaningful control over their personal data, they differ widely in how they define personal data and the mechanisms for exercising that control. 
For example, not all laws offer precise or comprehensive definitions of personal and sensitive data, nor do they align on the required granularity of consent. 
User rights also vary substantially. For example GDPR, while offering fine-grained information about personal, sensitive, and children data, does not include certain rights, such as the right to nominate, which is offered only under India's DPDPA. These variations influence how users understand their rights and how effectively they execute them.

Examining data protection laws through user-centric dimensions showed that GDPR and China's PIPL provide some of the strongest protections: each offers at least 8 out of 10  above-discussed rights to the user (see {\color{blue}Table~\ref{tab:privacy_rights}}) as well as requires specific consent (see {\color{blue}Table~\ref{tab:consent_granularity}}). 
In contrast, the CCPA adopts a flexible model, by not requiring prior consent for data collection, relies primarily on an opt-out mechanism, and provides only 7 out of 10 rights. 
As such, while user autonomy is a common objective across data protection laws, the level and form of protection differ significantly.

\bgroup
\def\arraystretch{1.25}
\begin{table*}[!h]
\scriptsize 
\centering
\begin{tabular}{|p{.5cm}|p{1.9cm}|p{1.2cm}|p{4.5cm}|p{1.5cm}|p{3cm}|p{1.4cm}|p{1.3cm}|}
\hline
\textbf{Law} & 
\textbf{Breach Notification Trigger} & 
\textbf{Notification deadline to authority} & 
\textbf{What to inform to the authority} &
\textbf{Notification deadline to users} & 
\textbf{What to inform to users} &
\textbf{Security Measures Required} & \textbf{Penalties} \\
\hline

GDPR & Personal data breach, 
Art.~34(1) 
& 72 hours after becoming aware of the breach (Art.~33) & 
(a) the categories and approximate number of individuals concerned, (b) the name and contact details of the data protection officer, the likely consequences of the personal data breach, the measures taken or proposed to be taken by the controller to address the personal data breach& 
Without undue delay, if high risk Art.~34(1) & 
b point of 4th column  Art.~34(2) & Techniques to achieve confidentiality, integrity, availability, Art.~32(1b)
& Up to €10m or 2\% of turnover Art.~83(4a)\\
\hline

CCPA & Breach of unencrypted personal information  (Art.~1798.150 and 1798.82) & 
``In the most expedient time possible'' (Art.~1798.82) & 
``What Happened, What Information Was Involved, What We Are Doing, What You Can Do, and For More Information'' Art.~1798.82(d) & 
Not explicit & Same as 4th column &  
Reasonable security procedures (Art.~1798.150) & 
\$100--\$750 per consumer per incident Art.~1798.150\,(a) (1) (A) \\
\hline

DPDPA & ``Personal data breach'' defined broadly Art.~2(u), 8(6)
& No time duration  
& \multicolumn{3}{c|}{Not explicit} 
& Reasonable security safeguards, Art.~8(5) 
& Up to rupee \INR{250} crore, Art.~33(2)
\\\hline

PIPL & Leak, falsification, or lost of personal information (Art.~57) & Immediate (Art.~57) & 
Types and causes of personal information leakage, falsification, and loss that have occurred or may occur, and the possible harm caused;
remedial measures taken by the organization and  individuals; contact information of the organization, Art.~57
& If measures to effectively avoid harm are not enough, Art.~57 
& 4th column 
& Technical measures, encryption, de-identification, Art.~51(3) 
& \multirow{3}{*}{\parbox[t]{1cm}{\raggedright Not explicit for data breaches}} \\ \cline{1-7}

LGPD & A security incident that may create risk or relevant damage to the individuals, Art.~48 & 
``In reasonable time,'' Art.~48(1) & 
Nature of the affected personal data, information on the individuals involved, an indication of the technical and security measures used to protect the data, and the risks related to the incident, the measures that were or will be adopted to reverse or mitigate the effects, Art.~48
& Reasonable time period, Art.~48(1) 
& Same as 4th column, Art.~48 
& Security, technical and administrative measures, Art.~46 
& \\ \cline{1-7}

POPIA & Security compromise of personal information, Art.~22 & 
``As soon as reasonably possible,''  Art.~22(2) & 
Description of possible consequences, measures taken or proposed, responsible party contacts 
& ``As soon as reasonably possible,'' Art.~ 22(2) 
& Possible consequences of the security compromise, measures to be taken, measures to be taken by the individual to mitigate the possible adverse effects, the identity of the unauthorised person (if known) who may have accessed or acquired the personal information, Art.~ 22(5) 
& Yes, Sec. 19  
& \\ \hline

\end{tabular}
\caption{Data breaches across data protection laws.}
\label{tab:breach}
\BBB\BBB\BBB
\end{table*}\egroup

\subsection{Laws in terms of Organizations}
\label{subsec:Laws in terms of Organizations}
\BB

\textbf{Objectives.} 
Organizations that collect and process personal, sensitive, and children's data are bound not only by the requirements of consent and user rights but also by a set of broader obligations designed to ensure lawful handling of data. These responsibilities define how organizations must operate when processing information, responding to incidents, transferring data across different jurisdictions, and demonstrating compliance to oversight authorities. 

In this section, our objective is to compare different data protection laws on the aspects of core processing principles to be followed by organizations ({\color{blue}\S\ref{subsubsec:Core Processing Principles}}), rules to be followed after data breaches ({\color{blue}\S\ref{subsubsec:Data Breaches}}), 
cross-jurisdictional data transfer rules ({\color{blue}\S\ref{subsubsec:Cross Jurisdiction Data Transfer}}), rules around data deletion ({\color{blue}\S\ref{subsubsec:Data Deletion}}), and 
 mechanisms to ensure accountability in following/implementing regulations ({\color{blue}\S\ref{subsubsec:Accountability}}).

\subsubsection{\bf Core Processing Principles}
\label{subsubsec:Core Processing Principles}

Data protection laws provide a set of foundational principles detailing how organizations should handle data to ensure they do not collect more than is required and securely store it. As will become clear from the definition of these principles, without them, user rights (such as consent withdrawal or access) cannot be meaningfully exercised or verified. For instance, consent will become meaningless without principles of purpose limitation, transparency, and data minimization. 

Below, we provide a superset of all the principles that are given in the data protection laws (note that the names of principles are directly taken from the laws): 

\begin{itemize}[leftmargin=0.01in]
\item
\noindent
\emph{\textbf{Purpose limitation}} (\textit{i}.\textit{e}., data must be collected and processed only for specific, explicit, and legitimate purposes), 

\item\noindent
\emph{\textbf{Fairness}} (\textit{i}.\textit{e}., processing does not harm individuals),

\item\noindent
\emph{\textbf{Transparency}} (\textit{i}.\textit{e}., inform individuals clearly about processing activities, purposes, and their rights), 

\item\noindent
\emph{\textbf{Data minimization}} (\textit{i}.\textit{e}., should collect only the minimum amount of personal data necessary to achieve the stated purpose).

\item\noindent
\emph{\textbf{Accuracy}} (\textit{i}.\textit{e}., personal data is accurate, complete, and up to date).

\item\noindent
\emph{\textbf{Storage limitation}} (\textit{i}.\textit{e}., personal data should not be kept longer than necessary for the purposes for which it was collected). 

\item\noindent
\emph{\textbf{Security}} (\textit{i}.\textit{e}., ensure data integrity and data confidentiality). 

\item\noindent
\emph{\textbf{Accountability}} (\textit{i}.\textit{e}., organizations demonstrating compliance with the principles). 

\end{itemize}
{\color{blue}Table~\ref{tab:core-principles}} compares each law under these principles. 
GDPR and Brazil's LGPD are the only two regulations that satisfy all the above-mentioned principles. 
In contrast, California's CCPA is the most flexible law that does not explicitly offer most of the principles. Furthermore, some laws do not directly offer such principles. For example, DPDPA does not explicitly offer data minimization; however, Article~6 requires consent-based specific processing, which turns into data minimization. 
Likewise, DPDPA does not explicitly mention storage limitations and accountability. However, specific data collection, as outlined in Article~8(8), and the organization's obligations, as outlined in Article~8, ensure both.  

\bgroup
\def\arraystretch{1.25}
\begin{table*}[!h]
\BBB
\scriptsize 
\centering
\begin{tabular}{|p{0.5cm}|p{3.1cm}|p{2cm}|p{2.9cm}|p{7cm}|p{2cm}|p{2.5cm}|}
\hline
\textbf{Law} & 
\textbf{Transfer Conditions} & 
\textbf{Review Mechanisms} & 
\textbf{Role of Individual's Consent} &
\textbf{Exemptions} \\
\hline

GDPR & Ensures adequate protection, Art.~45(1) 
& 
At least every four years, Art.~45(3) & 
Yes, Art.~49(1)(a)& 
Important reasons of public interest, the establishment, exercise or defense of legal claims, vital interests of individuals, Art.~49(1)(d,e,f)\\
\hline

CCPA &  \multicolumn{4}{c|}{Not applicable}
\\
\hline

DPDPA & By default, data transfer is allowed. The Government may restrict transfers by notification, Art.~16 & 
No & 
Consent alone is not sufficient unless the government permits & 
 Legal rights, the interest of prevention, detection, investigation, or prosecution of any offense, ascertaining the financial information and assets and liabilities of any person who has defaulted in payment,
Art.~17(1)(c,d,f) \\
\hline

PIPL & Only if approved by the State cyberspace administration, Art.~38& 
\multicolumn{3}{c|}{Not exists}
\\
\hline

LGPD& Ensures adequate protection, Art.~33(1) & 
Yes, but not duration bound, Art.~34 & 
Explicit consent for data transfer, Art.~33(8) & 
For international legal cooperation, protect life/ physical safety,
Art.~33(3,4) \\
\hline

POPIA  & Ensures an adequate level of protection, Art.~72(1a) & 
No & 
 Yes, Art.~72(1b) & 
the implementation of pre-contractual measures, the performance of a contract, and the benefit of the individual, Art.~72(1)(c,d,e)\\
\hline

\end{tabular}
\caption{Cross-jurisdiction data transfer across data protection laws.}
\label{tab:transfer}
\BBB\BBB
\end{table*}\egroup

\subsubsection{\bf Data Breaches}
\label{subsubsec:Data Breaches}
All data protection laws impose obligations on organizations to implement preventive 
and corrective measures to reduce the risk of data breaches, typically specifying:
(\textit{i})~the conditions triggering breach notification to regulators/authorities, affected individuals, or both,
(\textit{ii})~timelines for reporting to regulators/authorities and individuals,
(\textit{iii})~information to be disclosed to authorities and individuals regarding the breach,
(\textit{iv})~security measures to be implemented to prevent breaches, and
(\textit{v})~penalties applicable in the event of a breach.


{\color{blue}Table~\ref{tab:breach}} compares different laws on parameters like notification triggers, reporting timelines, required content of notices, and penalties.
Not all laws impose stringent conditions on what could trigger the organization to notify about data breaches. 
However, once organizations detect data breaches, only GDPR imposes a strict requirement to inform the regulator within 72 hours. 
In contrast, China's PIPL and Brazil's LGPD require organizations to provide fine-grained information about incidents to both users and regulators. 
However, India's DPDPA does not provide any explicit conditions on all such. 
Furthermore, China's PIPL, Brazil's LGPD, and South Africa's POPIA do not impose any specific penalties for data breaches. 
Only GDPR, California's CCPA, and India's DPDPA placed penalties for data breaches. 

\subsubsection{\bf Cross Jurisdiction Data Transfer}
\label{subsubsec:Cross Jurisdiction Data Transfer}

While data mobility enables global commerce and collaboration, it also jeopardizes individual rights and the potential loss of regulatory oversight once data leaves its country of origin.

To mitigate these risks, data protection laws impose conditions on cross-border data transfers to ensure that protection standards accompany the 
data. These provisions typically address four key aspects:
(\textit{i})~transfer conditions, which define the circumstances under which cross-border transfers are permitted;
(\textit{ii})~review mechanisms, which establish procedures for evaluating and approving or disallowing transfers;
(\textit{iii})~the role of individual consent, which in some jurisdictions requires explicit authorization from individuals even where the recipient country lacks adequate protection; and
(\textit{iv})~exemptions, which identify circumstances where transfers are permitted regardless of general restrictions.

{\color{blue}Table~\ref{tab:transfer}} compares these laws across the above-mentioned four aspects.
Note that India's DPDPA and China's PIPL require government approval to allow data transfers. Only GDPR allows cross-jurisdictional data transfers based on an individual's consent for data transfer, and it imposes a strict 4-year review deadline. In contrast, the CCPA is the only law that does not address cross-jurisdictional data transfers. 

\bgroup
\def\arraystretch{1.25}
\begin{table*}[h!]
\scriptsize 
\centering
\begin{tabular}{|p{0.5cm}|p{8cm}|p{3.8cm}|p{0.8cm}|p{2.4cm}|p{3cm}|}
\hline
\textbf{Law} & 
{\bf Reasons for Automatic Deletion} & 
\textbf{Exemptions} & \textbf{Right to Delete} & \textbf{Duration of deletion after executing right to delete} \\
\hline
GDPR & 

No longer than is necessary for the purposes for which the personal data are processed, Art.~5(1e)
& 
Public interest, scientific or historical research purposes or statistical purposes, Art.~5(1e)
& 
Yes, Art.~17 &
Without undue delay, Art.~17(1) \\
\hline

CCPA &  
Complete the transaction, security incidents, protect against malicious, deceptive, fraudulent, or illegal activity, debug to identify and repair errors that impair existing intended functionality, exercise free speech, ensure the right of another consumer to exercise that consumer's right of free speech, or exercise another right provided for by law, and comply with the California Electronic Communications Privacy Act, comply with a legal obligation, and purposes like scientific, historical, or statistical research, Art.~1798.105(d) &
Yes, Art.~1798.105(a) & Not given  & Not given
\\\hline

DPDPA  & 
Withdrawing consent or purpose achieved, Art.~8(7)(a) &
Unless retention of the same is necessary for the
specified purpose or  compliance with any law, Art.~13(3) &
Yes, Art.~12(3) &
Not given
\\\hline

PIPL & 
Fulfilled purpose -- Art.~19, 47(1), 
the agreed storage period has expired -- Art.~47(2), consent withdrawal -- Art.~47(3),
violation of laws by the organization -- Art.~47(4) & 
Administrative regulations, Art.~19
& Yes, Art.~47 & Not given\\
\hline

LGPD &  
Purpose achieved,
end of the processing period, right to revoke consent, a violation of the provisions, Art.~15 & 
 Legal obligation, research entity, Art.~16
 & Yes, Art.~18(6) & Not given
 \\\hline

POPIA& 
Purpose achieved, \textit{i}.\textit{e}., no longer authorized to retain the record, Art.~14(4) & 
Required or authorized by law, historical, statistical or research purposes, Art.~14(1),14(2)
& Yes, Art.~24(1) & As soon as possible, Art.~24(2)
\\\hline

\end{tabular}
\caption{Data deletion rules across data protection laws.}
\label{tab:retention}
\BBB\BBB
\end{table*}\egroup

\bgroup
\def\arraystretch{1.25}
\begin{table*}[h!]
\B
\scriptsize
\centering
\begin{tabular}{|p{0.5cm}|p{1cm}|p{2cm}|p{4cm}|p{3.8cm}|p{3.5cm}|}
\hline
{Law} & 
{Records of Processing} & 
{Data Protection Officer} &
{Impact Assessments} & 
{Regulatory Reporting Obligations} &
{Maximum Penalties} \\
\hline

GDPR & 
Mandatory, Art.~30 & 
Mandatory, only if sensitive and personal information collected at large scale, Art.~37 &
New technologies, high risk to the rights and freedoms of natural persons, processing on a large scale of sensitive and personal information, Art.~35(1) & 
Supervisory authorities may ask for records of processing, Art.~30(4) & 
Up to €20M or 4\% of global annual turnover, Art.~83
\\\hline

CCPA & 
\multicolumn{4}{c|}{Not defined for keeping the processing records} & 
  \$7,500 per intentional violation, \$2,500 per unintentional violation, Art.~1798.155 \\
\hline

DPDPA & Not defined & 
Mandatory, Art.~10(2) & 
Periodic assessment report should include a description of the rights of individuals, the purpose of
processing of their personal data, the risk to the rights of individuals,  Art.~10(3) & 
Ensure compliance, Clause 27& 
Rupee \INR{250}, Art.~33 \\\hline

PIPL & 
Mandatory, Art.~55 & 
Mandatory, Art.~52 & 
Impact assessment will be done, but what it will include is not defined, Art.~55 & 
State cyberspace administration is deeply involved, Art.~60-65 & 
Up to RMB 50M or 5\% of annual turnover, Art.~66 \\
\hline

LGPD & 
Mandatory, 37 &
Yes, Art.~41 & 
Should include the description concerning the personal data processing that could pose risks to civil liberties and fundamental rights, as well as measures, safeguards
and mechanisms to mitigate said risk --- Art.~5(17) & 
The national authority may determine that the controller must prepare a data
protection impact assessment, which shall include personal data, sensitive data, and  data processing operations, Art.~38  & 
Up to 2\% of revenue in Brazil, capped at 50M BRL per violation, Art.~52(2) \\
\hline

POPIA & 
Mandatory, Art.~17 &
Mandatory, Art.~55 & 
Not defined & 
Providing authorization to organizations before they collect personal data, Art.~57(1) & 
At most 10M  or imprisonment up to 10 years, Art.~107 and 109(2)(c)\\\hline

\end{tabular}
\caption{Accountability rules across data protection laws.}
\label{tab:accountability}
\BBB\BBB\BB
\end{table*}\egroup

\subsubsection{\bf Data Deletion}
\label{subsubsec:Data Deletion}
A core principle of data protection law is that personal data must not be stored indefinitely. Organizations are required to adopt retention policies ensuring data is kept only as long as necessary to fulfill the purpose for which it was collected. Once that purpose is achieved, or the legal basis for processing ceases to exist, organizations must securely delete, anonymize, or otherwise render the data unusable.

{\color{blue}Table~\ref{tab:retention}} compares the requirements across different data protection laws with respect to data deletion on parameters, such as:
(\textit{i})~the circumstances under which organizations must automatically delete data without individual intervention;
(\textit{ii})~the exemptions that allow continued retention even when individuals request deletion;
(\textit{iii})~the existence of a formal right to deletion (or erasure); and
(\textit{iv})~the time frame within which organizations must comply with a deletion request once the right is exercised. 

Note that while all laws require data deletion once the purpose is fulfilled, only GDPR and POPIA further mandate deletion without undue delay. Others do not specify a timeline for deleting data after the purpose is completed. Furthermore, all laws, except California's CCPA, give individuals the right to delete.

\bgroup
\def\arraystretch{1.25}
\begin{table*}[h!]
\scriptsize
\centering
\begin{tabular}{|p{0.5cm}|p{3.7cm}|p{3.6cm}|p{5cm}|p{2.5cm}|}
\hline
\textbf{Law} & \textbf{Regulatory Authority} & {Independence} & \textbf{Appointment} & \textbf{Term Duration} \\\hline

GDPR & Supervisory Authorities in each EU member state, Art.~51(1)   & 
Yes, Art.~52(1) & Appointed nationally according to national law, Art.~53 and 54  & At least 4 years, Art.~54(1)(d)\\\hline

CCPA & 
California Privacy Protection Agency, Art.~1798.199.10 & 
Not explicitly mentioned & 
Appointed by the Governor, Art.~1798.199.10 & 
At most 8 years, Art.~1798.199.20 \\\hline

DPDPA & 
Data Protection Board of India, Art.~18(1)  &
Yes, Art.~28(1)   & 
The Indian Government, Art.~19(2)  & 
2 years, Art.~20(2)  \\\hline

PIPL & 
 State cyberspace administration, Art.~60   &
 Not independent, work under relevant departments under the State Council, Art.~60  & 
 Not given & 
 Not given \\
\hline

LGPD & 
National Data Protection Authority (ANPD), Art.~55(1)   & 
Yes, Art.~55(2)   & 
Appointed by the President of Brazil and approval by the Federal Senate, Art.~55(4)(1)    &
4 years, Art.~55(4)(3)  \\\hline

POPIA & 
Information Regulator, Art.~39   & 
Yes, Art.~39(2) & 
Members appointed by President President, Art.~41(2)(a)  & 
At most 5 years, Art.~41(3)  \\\hline

\end{tabular}
\caption{Regulatory authorities across data protection laws.}
\label{tab:authorities}
\BBB\BBB\BBB
\end{table*}\egroup

\subsubsection{\bf Accountability}
\label{subsubsec:Accountability}

Accountability requires organizations not only to comply with data protection obligations but also to \textit{demonstrate compliance in a verifiable manner}. This principle ensures that organizations maintain internal governance, monitor risks, and provide transparency to regulators and individuals. 

{\color{blue}Table~\ref{tab:accountability}} compares the accountability principles across different parameters, such as:  
(\textit{i})~\textit{Records of Processing}, which captures the obligation of organizations to maintain detailed information about the personal data they process and the purposes of processing;  
(\textit{ii})~\textit{Data Protection Officer}, a designated officer responsible for overseeing compliance, implementing data protection policies, and serving as the point of contact with regulators and individuals;  
(\textit{iii})~\textit{Impact Assessments}, reflecting whether privacy or risk assessments must be conducted prior to high-risk processing, such as large-scale collection of sensitive data or use of new technologies;  
(\textit{iv})~\textit{Regulatory Reporting Obligations}, which describe the requirements for organizations to provide reports to government authorities to demonstrate compliance; and  
(\textit{v})~\textit{Maximum Penalties}.

Except for India's DPDPA, all laws require organizations to retain records for processing. However, all laws require an organization to appoint a data protection officer. China's PIPL is the only law that requires a higher authority, namely the State Cyberspace Administration, to be deeply involved in the accounting process. 
In contrast, California's CCPA is the only law that does not impose any accountability rule for the keeping the record of processing. However, CCPA requires maintaining records of other activities, such as 
request by a consumer to opt out of the sale,  business compliance with a consumer's opt-out request, Article~1798.185(a)(4).

\subsubsection{\bf Lessons}
\label{subsubsec:lesson organizations}

Data protection laws share common foundational principles, such as lawful processing, transparency, purpose limitation, data minimization, and data security, yet the rigor with which these obligations are implemented and enforced varies considerably across jurisdictions.
Among these obligations, accountability plays the most critical role in building user confidence: when organizations are required to document their practices, justify their processing activities, and demonstrate compliance, other obligations, such as breach management, cross-border transfer controls, and data deletion, naturally become more enforceable and verifiable.

In terms of accountability, GDPR and PIPL impose the most demanding organizational requirements, including strict purpose limitation and documentation of processing. On the other hand, CCPA follows a more flexible, business-oriented model with fewer procedural burdens. DPDPA, LGPD, and POPIA fall between these extremes: they introduce meaningful organizational responsibilities but allow jurisdiction-specific exceptions and flexibilities that limit uniformity in operational practice.

\bgroup
\def\arraystretch{1.25}
\begin{table*}[h]
\centering
\scriptsize
\begin{tabular}{|p{0.5cm}|p{5.5cm}|p{7cm}|p{3.25cm}|}
\hline
\textbf{Law} & \textbf{Investigation Triggers} & \textbf{Investigative Powers} &
\textbf{Corrective Power} \\ \hline

GDPR &
Complaints lodged by an individual, another supervisory authority, or other public authority
Art.~57(1)(f), 57(1)(h)
&
Audits and 
review
Art.~58(1)(b), 58(1)(c)
&
Issue warnings, 
issue reprimands,
a temporary or definitive limitation, including a ban on processing, Art.~58(2)
\\ \hline

CCPA  &
California Privacy Protection Agency-initiated Art.~1798.199.55.(a) and complaints,  Art.~1798.199.45
&
Subpoena witnesses, compel their attendance and testimony, administer oaths and affirmations, take evidence and require by subpoena the production of any books, papers, records, or other items material to the performance of the agency's duties or exercise of its powers, including, but not limited to, its power to audit a business’ compliance with this title, Art.~1798.199.65. & Not given 
 \\ \hline

DPDPA & 
An intimation of personal data breach, and a complaint made by an individual regarding a personal data breach, Art.~27(1)
&
An inquiry, summoning and inspecting any data, book, document, register, books of account or any
other document, Art.~28(3), 28(7) &
Not given
   \\ \hline

PIPL   &
Complaints and reports relating to personal information protection, and illegal personal information processing activities, Art.~61(2), 61(4)
&
An inquiry, 
on-site inspection, and  sealing up or seizing the equipment, Art.~63
&Not given 
\\ \hline

LGPD &
Pleadings from an individual against the organization after the individual has demonstrated that he/she presented a complaint against the organization that was not solved in the timeframe established in the law, Art.~55J(5) 
&
Audit, Art.~55J(16) and
apply sanctions,  Art.~55J(4), 55K
& Not given
 \\ \hline

POPIA &
Complaints from individuals, Art.~74
&
Pre-investigation.
Art.~76(1)a, 79,
summon, Art.~81(1)a,
search any premises, Art.~81(1)d, 
and a warrant by the High Court, a regional magistrate or a magistrate, Art.~82 &
Not given\\ \hline

\end{tabular}
\caption{Investigations under data protection laws. }
\label{tab:supervision}
\BBB\BBB\BBB
\end{table*}\egroup

\B
\subsection{Laws in terms of Government}
\label{subsec:Laws in terms of Government}
\BB

\textbf{Objectives.} 
The effectiveness of data protection laws depends not only on organizational compliance but also on the oversight provided by government and regulatory authorities. While accountability obligations require organizations to implement internal mechanisms for data protection, these measures must be complemented by independent authorities empowered to supervise, investigate, and enforce compliance. Data protection laws, therefore, establish regulatory bodies with varying levels of independence, appointment mechanisms, and operational mandates. 

Our objective in this section is to analyze the role of governments and regulators under different data protection laws.
{\color{blue}\S\ref{subsubsec:Regulatory Authorities}} will discuss how different laws establish regulatory
bodies, and {\color{blue}\S\ref{subsubsec:Investigations by the Authorities}} will discuss how regulatory
bodies can investigate organizations in their jurisdictions.

\subsubsection{\bf Regulatory Authorities}
\label{subsubsec:Regulatory Authorities}
Regulatory authorities are responsible for ensuring the effective implementation and enforcement of data protection laws, overseeing compliance, investigating complaints, issuing guidance, and imposing corrective measures. 

{\color{blue}Table~\ref{tab:authorities}} compares the regulatory authorities in terms of four parameters:
(\textit{i})~\textit{Regulatory Authority}, lists the designated body responsible for law enforcement;
(\textit{ii})~\textit{Independence}, indicates whether the authority operates independently or is subordinate to government entities;
(\textit{iii})~\textit{Appointment}, describes how the members of the authority are appointed; and
(\textit{iv})~\textit{Term Duration}, specifies the length of service for the members of the authority.

In summary, 
while all data protection laws formally establish regulatory authorities, their design and autonomy differ substantially. GDPR, LGPD, and POPIA rely on autonomous regulators with defined appointment procedures and fixed terms. In contrast, China's PIPL places regulatory oversight under the State Cyberspace Administration, resulting in a highly centralized model with limited structural independence and no specified term durations. DPDPA and CCPA fall between these extremes.

\subsubsection{\bf Investigations by the Authorities}
\label{subsubsec:Investigations by the Authorities}

Government authorities are vested with distinct investigative powers that determine how they monitor compliance, initiate inquiries, and respond to violations of data protection laws.
{\color{blue}Table~\ref{tab:supervision}} compares these laws on the following three dimensions:
(\textit{i})~\emph{Investigation triggers} that specify the conditions under which regulators are authorized to initiate investigations, such as complaints from individuals, breach notifications, or regulator-initiated inquiries;
(\textit{ii})~\emph{Investigative powers} that define the procedural authorities granted to regulators to examine compliance, including conducting audits, summoning individuals, demanding documents and records, performing on-site inspections, and, in some jurisdictions, seizing equipment or individuals to judicial authorization; and
(\textit{iii})~\emph{Corrective powers} that describe measures available to regulators to remedy non-compliance.

In summary, GDPR offers corrective powers.
DPDPA and POPIA offer strong investigative authority. Under DPDPA, the Data Protection Board is empowered to exercise powers similar to those of a civil court of India. Similarly, under POPIA, the Information Regulator may conduct investigations and seek court-issued warrants, with judicial involvement.

\section{Conclusion}
\label{sec:conclusion}
\BB
This paper presented a comparative analysis of privacy regulations that cover approximately 38.3\% of the world's population.
We examined these laws through a data life-cycle lens to understand how user rights, organizational obligations, and government enforcement mechanisms are defined and executed across the phases of data collection, storage, processing, sharing, deletion, and bequeathal.
Our analysis shows that, although data protection laws pursue similar high-level goals, their interpretations, scope, and enforcement vary substantially across jurisdictions and across different stages of the data life cycle.

We further discussed that many practical compliance challenges faced by both organizations and individuals stem from non-uniform definitions and vague operational requirements in privacy regulations. 
Consequently, organizations struggle to design end-to-end compliant systems that function consistently across regulatory boundaries, while individuals often lack effective means to exercise and verify their rights beyond the data collection phase.
Our analysis highlights the need for systematic end-to-end frameworks that bridge legal requirements with technical system design.

\BB
\section*{Acknowledgments}
\BB
{
Ritwik Banerjee was supported in part by the AI Innovation Institute at Stony Brook University (Award No. 102316; Project  No. 1194042) and the U.S. National Science Foundation (Award No. CNS 2335686). 
Work of Indrakshi Ray and Ethan Myers was supported by NSF Award Number CNS 2335687 and State of Colorado Cybersecurity funding under SB 18-086.}


\begin{thebibliography}{10}

\bibitem{IDP}
Interdependent Privacy (IDP): \url{tinyurl.com/4n9p9d3p}.

\bibitem{DLA}
DLA Piper: \url{www.dlapiperdataprotection.com}.

\bibitem{onetrust}
OneTrust DataGuidance (IDP): \url{tinyurl.com/yc5vvfpf}.

\bibitem{10.5555/3563572.3563580}
D.~Abrokwa et~al.
\newblock Comparing security and privacy attitudes among {U.S.} users of different smartphone and smart-speaker platforms.
\newblock In {\em {SOUPS}}, pages 139--158, 2021.

\bibitem{ardalani2024}
A.~Ardalani et~al.
\newblock Towards precise detection of personal information leaks in mobile health apps.
\newblock {\em CoRR}, abs/2410.00277, 2024.

\bibitem{10.5555/3361476.3361480}
O.~Ayalon et~al.
\newblock Evaluating users' perceptions about a system's privacy: differentiating social and institutional aspects.
\newblock In {\em {SOUPS}}, page 41–59, 2019.

\bibitem{DBLP:conf/sp/BarthDMN06}
A.~Barth et~al.
\newblock Privacy and contextual integrity: Framework and applications.
\newblock In {\em {IEEE SP}}, pages 184--198, 2006.

\bibitem{Boutet_Morel_2025}
A.~Boutet et~al.
\newblock {``I'm} not for sale" - perceptions and limited awareness of privacy risks by digital natives about location data.
\newblock In {\em {AAAI}}, pages 277--288, 2025.

\bibitem{Dhondt_294510_2024}
K.~Dhondt et~al.
\newblock Swipe left for identity theft: An analysis of user data privacy risks on location-based dating apps.
\newblock In {\em {USENIX} Security}, pages 5053--5070, 2024.

\bibitem{fan_empirical_2020}
M.~Fan et~al.
\newblock An empirical evaluation of {GDPR} compliance violations in {Android} {mHealth} apps.
\newblock In {\em {ISSRE}}, pages 253--264, 2020.

\bibitem{feichtner_2020}
J.~Feichtner et~al.
\newblock Understanding privacy awareness in {Android} app descriptions using deep learning.
\newblock In {\em {CODASPY}}, pages 203--214, 2020.

\bibitem{feng_acnet_2019}
Y.~Feng et~al.
\newblock {AC-Net}: Assessing the consistency of description and permission in android apps.
\newblock {\em IEEE Access}, 7:57829--57842, 2019.

\bibitem{ferreira_rulekeeper_2023}
M.~Ferreira et~al.
\newblock {RuleKeeper}: {GDPR}-aware personal data compliance for web frameworks.
\newblock In {\em {IEEE SP}}, pages 2817--2834, 2023.

\bibitem{herwanto_leveraging_2024}
G.~B. Herwanto et~al.
\newblock Leveraging {NLP} techniques for privacy requirements engineering in user stories.
\newblock {\em IEEE Access}, 12:22167--22189, 2024.

\bibitem{InterdependentPrivacy}
M.~Humbert et~al.
\newblock A survey on interdependent privacy.
\newblock {\em ACM Comput. Surv.}, 52(6), 2019.

\bibitem{jia_who_2019}
Q.~Jia et~al.
\newblock Who leaks my privacy: Towards automatic and association detection with {GDPR} compliance.
\newblock In {\em {WASA}}, pages 137--148, 2019.

\bibitem{koch_ok_nodate}
S.~Koch et~al.
\newblock The {OK} is not enough: {A} large scale study of consent dialogs in smartphone applications.
\newblock In {\em {USENIX} Security}, pages 5467--5484, 2023.

\bibitem{kollnig_fait_nodate}
K.~Kollnig et~al.
\newblock A fait accompli? an empirical study into the absence of consent to third-party tracking in android apps.
\newblock In {\em {SOUPS}}, 2021.

\bibitem{kramer_death_2024}
J.~Krämer.
\newblock The death of privacy policies: {How} app stores shape {GDPR} compliance of apps.
\newblock {\em Internet Policy Review}, 13(2), 2024.

\bibitem{li_are_2024}
S.~Li et~al.
\newblock Are we getting well-informed? an in-depth study of runtime privacy notice practice in mobile apps.
\newblock In {\em {CCS}}, pages 1581--1595, 2024.

\bibitem{ihunter_2024}
D.~Liu et~al.
\newblock ihunter: Hunting privacy violations at scale in the software supply chain on {iOS}.
\newblock In {\em {USENIX} Security}, 2024.

\bibitem{liu_opted_2024}
Z.~Liu et~al.
\newblock Opted out, yet tracked: Are regulations enough to protect your privacy?
\newblock {\em {POPETs}}, 2024:280--299, 01 2024.

\bibitem{mohan_analyzing_2019}
J.~Mohan et~al.
\newblock Analyzing {GDPR} compliance through the lens of privacy policy.
\newblock In {\em Heterogeneous Data Management, Polystores, and Analytics for Healthcare}, page 82–95, 2022.

\bibitem{abhinav_2021}
A.~Mohanty et~al.
\newblock Hybridiagnostics: Evaluating security issues in hybrid smarthome companion apps.
\newblock In {\em IEEE Security and Privacy Workshops}, pages 228--234, 2021.

\bibitem{neptune_2022}
S.~Neupane et~al.
\newblock On the data privacy, security, and risk postures of iot mobile companion apps.
\newblock In {\em {DBSec}}, page 162–182, 2022.

\bibitem{nguyen_share_nodate}
T.~T. Nguyen et~al.
\newblock Share first, ask later (or never?) studying violations of {GDPR{\textquoteright}s} explicit consent in android apps.
\newblock In {\em {USENIX} Security}, page 3667–3684, 2021.

\bibitem{nguyen_freely_2022}
T.~T. Nguyen et~al.
\newblock Freely given consent? studying consent notice of third-party tracking and its violations of {GDPR} in {Android} apps.
\newblock In {\em {CCS}}, page 2369–2383, 2022.

\bibitem{nissenbaum2004privacy}
H.~Nissenbaum.
\newblock Privacy as contextual integrity.
\newblock {\em Wash. L. Rev.}, 79:119, 2004.

\bibitem{pandita_whyper_2013}
R.~Pandita et~al.
\newblock {WHYPER}: towards automating risk assessment of mobile applications.
\newblock In {\em {USENIX} Security}, page 527–542, 2013.

\bibitem{qamar_detecting_2021}
A.~Qamar et~al.
\newblock Detecting compliance of privacy policies with data protection laws.
\newblock {\em CoRR}, abs/2102.12362, 2021.

\bibitem{qu_autocog_2014}
Z.~Qu et~al.
\newblock Autocog: Measuring the description-to-permission fidelity in android applications.
\newblock In {\em {CCS}}, page 1354–1365, 2014.

\bibitem{Reyes2017IsOC}
I.~Reyes et~al.
\newblock ``{Is} our children's apps learning?" automatically detecting coppa violations.
\newblock In {\em Workshop on Technology and Consumer Protection}, 2017.

\bibitem{reyes_wont_2018}
I.~Reyes et~al.
\newblock {``Won't} somebody think of the children?" examining {COPPA} compliance at scale.
\newblock {\em {POPETs}}, pages 63--83, 2018.

\bibitem{sassetti_assurance_2023}
G.~Sassetti et~al.
\newblock Assurance, consent and access control for privacy-aware {OIDC} deployments.
\newblock In {\em {DbSec}}, pages 203--222, 2023.

\bibitem{schmidt_2025}
D.~Schmidt et~al.
\newblock Leaky apps: Large-scale analysis of secrets distributed in android and ios apps.
\newblock In {\em {CCS}}, pages 2459--2473, 2025.

\bibitem{singh_technical_2020}
R.~G. Singh et~al.
\newblock A technical look at the indian personal data protection bill.
\newblock {\em CoRR}, abs/2005.13812, 2020.

\bibitem{xiang_policychecker_2023}
A.~Xiang et~al.
\newblock {PolicyChecker}: Analyzing the {GDPR} completeness of mobile apps' privacy policies.
\newblock In {\em {CCS}}, pages 3373--3387, 2023.

\bibitem{zhang-kennedy_whether_nodate}
L.~Zhang{-}Kennedy et~al.
\newblock {``Whether} it's moral is a whole other story": Consumer perspectives on privacy regulations and corporate data practices.
\newblock In {\em {SOUPS}}, pages 197--216, 2021.

\bibitem{zimmeck_automated_2017}
S.~Zimmeck et~al.
\newblock Automated analysis of privacy requirements for mobile apps.
\newblock In {\em {NDSS}}, 2017.

\bibitem{sebastian_zimmeck_maps_2019}
S.~Zimmeck et~al.
\newblock Maps: Scaling privacy compliance analysis to a million apps.
\newblock {\em {POPETs}}, 2019:66--86, 2019.

\bibitem{zufferey_revoked_2023}
N.~Zufferey et~al.
\newblock ``{Revoked} just now!" {Users'} behaviors toward fitness-data sharing with third-party applications.
\newblock {\em {POPETs}}, 2023:47--67, 2023.

\end{thebibliography}
\end{document}